\documentclass[twocolumn,showpacs,preprintnumbers,amsmath,amssymb]{revtex4}

\usepackage{graphicx,epsf}

\def\bv {\mathbf{v}}

\def\bb {\mathbf{b}}
\def\bz {\mathbf{z}}
\def\bu {\mathbf{u}}
\def\bx {\mathbf{x}}

\def\bw {\mathbf{w}}

\input epsf

\begin{document}

\title{Scaling laws of turbulence and heating of fast solar wind: the role of
density fluctuations}

\author{V. Carbone$^{1,2}$, R. Marino$^{1,3}$, L. Sorriso-Valvo$^2$,
A. Noullez$^3$, R. Bruno$^4$}

\affiliation{$^1$ Dipartimento di Fisica, Universit\`a della Calabria, Ponte Bucci 31C, I-87036 Rende (CS), Italy \\
$^2$ Liquid Crystal Laboratory, INFM/CNR, Ponte Bucci 33B,
I-87036 Rende (CS), Italy \\
$^3$ Observatoire de la C\^ote d'Azur, Boulevard de l'Observatoire,
Nice Cedex 04, France \\
$^4$ Istituto di Fisica dello Spazio Interplanetario --
INAF, via Fosso del Cavaliere Roma, Italy
}

\begin{abstract}

Incompressible and isotropic magnetohydrodynamic turbulence in plasmas can be
described by an exact relation for the energy flux through the scales. This
Yaglom-like scaling law has been recently observed in the solar wind above the
solar poles observed by the Ulysses spacecraft, where the turbulence is in an
Alfv\'enic state.  An analogous phenomenological scaling law, suitably modified 
to take into account compressible fluctuations, is observed more frequently in the 
same dataset.  Large scale density fluctuations, despite their low amplitude, play
thus a crucial role in the basic scaling properties of turbulence. The turbulent 
cascade rate in the compressive case can moreover supply the energy dissipation 
needed to account for the local heating of the non-adiabatic solar wind.

\end{abstract}


\date{\today}

\maketitle

The interplanetary space is permeated by the solar wind~\cite{generale}, a
magnetized, supersonic flow of charged particles originating in the high solar
atmosphere and blowing away from the sun. Low frequency fluctuations of solar
wind variables are often described in the framework of fully developed
hydromagnetic (MHD) turbulence~\cite{tu,noi}. The large range of scales
involved, spanning from~1\,AU ($\simeq 1.5 \times 10^8$\,km) down to a few
kilometers, make the solar wind the largest ``laboratory'' where
MHD~turbulence can be investigated using measurements collected {\it in situ\/}
by instruments onboard spacecraft~\cite{noi}.
MHD~turbulence is often investigated through the Els\"asser
variables~$\bz^{\pm} = \bv \pm (4\pi\rho)^{-1/2}\,\bb$, computed from the local
plasma velocity~$\bv$ and magnetic field~$\bb$, $\rho$ being the plasma mass
density. In terms of such variables, MHD~equations can be rewritten as
$\partial_t {\bf z}^\pm+{\bf z}^\mp \cdot {\bf \nabla}{\bf z}^\pm = -{\bf
\nabla} P/\rho + \mbox{\sl diss}$, where $P$ is the total hydromagnetic
pressure, and {\sl diss\/} indicates dissipative terms involving the
viscosity and the magnetic diffusivity.  As in the Navier-Stokes
equations for neutral fluids, the nonlinear terms~${\bf z}^\mp \cdot {\bf
\nabla}{\bf z}^\pm$ cause the turbulent energy transfer between
different scales, at high Reynolds numbers where dissipative terms can be
neglected.  However, in the MHD~case, they couple the two Els\"asser variables,
so that the Alfv\'enic MHD~fluctuations~${\bf z}^\pm$, propagating along the
background magnetic field, are advected by fluctuations~${\bf z}^\mp$
propagating in the opposite direction. The presence of strong correlations (or
anti-correlations) between velocity and magnetic fluctuations, along with a
nearly constant magnetic intensity and low amplitude density fluctuations, is
usually referred to as {\it Alfv\'enic state} of turbulence, and implies that one of the
two modes~${\bf z}^\pm$ should be negligible, making the nonlinear term of MHD
equations vanish for pure Alfv\'enic fluctuations. In that case, the turbulent
energy transfer should also disappear~\cite{dmv}.  
Alfv\'enic turbulence is observed almost ubiquitously in fast wind. This holds
both in the ecliptic fast streams, and in the high latitude wind blowing
directly from the sun coronal holes~\cite{belcher,bruno1,noi}. As pointed out
in~\cite{dmv}, the observation of Alfv\'enic state turbulence in the solar wind
represents therefore a paradoxical ``contradiction in terms''.

MHD~turbulence however satisfies an important {\em analytical\/} relation, which is
the equivalent for magnetized fluids of the Kolmogorov or the Yaglom relations.
Under suitable hypotheses, it has been shown~\cite{pp,noiprl} that the
pseudo-energy fluxes~$Y^{\pm}(\ell)$ through the scale~$\ell$ of the increments
of the Els\"asser fields $\Delta \bz^{\pm}(\ell) =
\bz^{\pm}(\bx+\ell)-\bz^{\pm}(\bx)$ follow a linear scaling relation
\begin{equation}
 Y^{\pm}(\ell) \equiv
 \langle |\Delta \bz^{\pm}|^2 \Delta z_\parallel^{\mp} \rangle =
 -\,\frac{4}{3} \,\epsilon^{\pm}\, \ell \ .
 \label{yaglom}
\end{equation}
Here, $\Delta z_\parallel^{\pm}$ represents the component of the
increment~$\Delta \bz^{\pm}$ along the direction~$\ell$, and $\epsilon^{\pm}$
are the dissipation rates per unit mass of the pseudo-energies $\langle
|\bz^{\pm}|^2 \rangle /2$ ($\langle\cdot\rangle$ indicates space averages). The
scaling law (\ref{yaglom}) has been recently observed experimentally in
polar wind~\cite{noiprl} and in the ecliptic plane~\cite{vasquez,mc}. The
confirmation of the scaling law~(\ref{yaglom}) is an important step towards the
solution of the apparent paradox of the Alfv\'enic turbulent state, because it
unambiguously shows that an MHD~cascade is present, maybe with a weak transfer
rate, despite the strong velocity-magnetic fields correlations.

Relation (\ref{yaglom}), which is of general validity within MHD turbulence, is
not always realized in the solar wind observations~\cite{noiprl}. 
Indeed, it is possible that local characteristics of the solar wind plasma do not 
always satisfy the assumptions required for~(\ref{yaglom}) to be valid, namely large-scale
homogeneity, isotropy, and incompressibility.  
The role of anisotropy of solar wind turbulence, 
which is expected to be important, is not considered in this letter. 
Density fluctuations in solar wind have a low amplitude, so that nearly 
incompressible MHD framework is usually considered~\cite{nearly,controversia}. 
However, compressible fluctuations are observed, typically convected structures
characterized by anti-correlation between kinetic pressure and magnetic
pressure~\cite{tm94}. Properties and interaction of the basic MHD modes in
the compressive case have also been considered in the past~\cite{goldreich,cho}. 
In the present paper, we show that density fluctuations,
despite their very low amplitude, play a central role in the turbulent
energy transfer, and that a phenomenological scaling law obtained by taking into account density
fluctuations is observed in a much larger proportion of fast solar wind. We also 
show that the turbulent dissipation can account for a large fraction of
the local heating causing a slower than expected decrease of temperature with
distance.  
A first attempt to include density fluctuations in the framework of fluid
turbulence was due to Lighthill~\cite{ligh55}. He pointed out that in a
compressible energy cascade, the mean energy transfer rate {\em per unit
volume\/}~$\epsilon_V \sim \rho v^3/\ell$ should be constant in a
statistical sense ($v$~being the characteristic velocity fluctuations at the
scale $\ell$), obtaining $v \sim (\ell/\rho)^{1/3}$. Fluctuations of a
density-weighted velocity field~$\bu \equiv \rho^{1/3} \bv$ should thus follow
the usual Kolmogorov scaling~$u^3 \sim \ell$. The same phenomenological conjecture can be
introduced in MHD~turbulence by considering the pseudo-energy dissipation rates
per unit volume $\epsilon_V^{\pm} \equiv \rho\epsilon^{\pm}$, and introducing
density-weighted Els\"asser~fields, defined as $\bw^{\pm} \equiv \rho^{1/3}
\bz^{\pm}$.  The equivalent of the Yaglom-type relation
\begin{equation}
 W^{\pm}(\ell) \equiv
 \langle |\Delta \bw^{\pm}|^2 \Delta w_\parallel^{\mp} \rangle
 \ \langle \rho \rangle^{-1} =
 -\,\frac{4}{3} \,\epsilon^{\pm}\, \ell
 \label{yagc}
\end{equation}
should then hold for the density-weighted increments~$\Delta
\bw^{\pm}(\ell)$.  Note that we have defined the flux~$W^{\pm}(\ell)$ so that
it reduces to~$Y^{\pm}(\ell)$ in the case of constant density, allowing for
comparisons between the compressible scaling~(\ref{yagc}) and the purely
incompressible one~(\ref{yaglom}).  Despite its simple phenomenological
derivation, the introduction of the density fluctuations in the Yaglom-type
scaling~(\ref{yagc}) seems to describe correctly the turbulent cascade for
compressible fluid (or magnetofluid) turbulence. The law for the velocity field
has been observed in recent numerical simulations~\cite{apj1,apj2}. 

We will now study the cascade properties of compressive MHD turbulence from solar
wind data collected by spacecraft Ulysses. In order to avoid as far as possible
variations due to solar activity, or other ecliptic disturbances such as slow 
wind sources, coronal mass ejection, current sheets, we
concentrate our analysis on pure Alfv\'enic state turbulence observed in high speed
polar wind. We use here measurements from Ulysses spacecraft in the first six 
months of~1996. Such period was characterized by low solar activity, so that solar 
origin disturbances werebv almost absent. 
Moreover, the spacecraft orbit was at high and slowly decreasing
heliolatitude, from about~$55\,{}^\circ$ to~$30\,{}^\circ$, and presented small
variations of the heliocentric distance~$r$, from 3\,AU to 4\,AU. Since the
mean wind speed $\langle \bv \rangle$ in the spacecraft frame is much larger
than the typical velocity fluctuations, and is nearly aligned with the radial
direction~$R$, space scales~$\ell$ can be viewed as time scales~$\tau$, related
through the Taylor hypothesis by $\ell = -\,\langle v_R \rangle\, \tau$. We
then used 8~minutes averaged time series of both Els\"asser
variables~$\bz^{\pm}(t)$ and density~$\rho(t) = n_{\rm p} + 4n_{\rm He}$
(obtained as the sum of proton density and 4~times He density), to compute the
density-weighted time series $\bw^{\pm}(t)$. From this time series we calculate
the increments $\Delta \bw^{\pm}(\tau) = \bw^{\pm}(t+\tau)-\bw^{\pm}(t)$ for
different time lags~$\tau$, and the third-order mixed structure functions
$W^{\pm}(\tau) = \langle |\Delta \bw^{\pm}(\tau)|^2 \Delta w^{\mp}_R(\tau)
\rangle_t$ by time averaging $\langle \cdot \rangle_t$ over windows of fixed
duration~$t$. The same procedure has been also been used to calculate the
quantities $Y^{\pm}(\tau)$ using the time series of the Els\"asser
fields~$\bz^{\pm}(t)$.  
In order to eliminate instationarities, heliolatitude
and heliocentric distance changes, and to explore the wind properties locally,
averages are computed over a moving window of about 11~days, consisting of
2048~data points. Accuracy of third order moments estimate~\cite{podesta1}
was tested with such sample size~\cite{dewit}.
We found that the third-order structure functions $W^{\pm}(\tau)$ computed from
the Ulysses data show a linear scaling
\begin{equation}
 W^{\pm}(\tau) \sim \frac{4}{3} \,\epsilon^{\pm}\, \langle v_R \rangle\, \tau
 \label{scal}
\end{equation}
during a considerable fraction of the period under study. In particular, we
observed linear scaling of $W^{+}(\tau)$ in about half of the signal, while
$W^{-}(\tau)$ displays scaling on about a quarter of the sample. 
As comparison, the corresponding incompressive scaling law for~$Y^{\pm}(\tau)$ 
was only observed in a third of the whole period, considerably smaller than the compressible
case~\cite{noiprl}. The portions of wind where the scaling is present are distributed 
in the whole period, and their extensions span from 6~hours up to 10~days. The linear scaling law
generally extends on about 2~decades, from a few minutes to one day or more.
For the compressible scaling, the two fluxes $W^{\pm}(\tau)$ coexist in a large number of cases. 
This does not hold for the incompressive scaling, where in general the scaling periods
for the two fluxes $Y^{\pm}(\tau)$ are disjoint.

\begin{figure}
 \includegraphics[width=8cm]{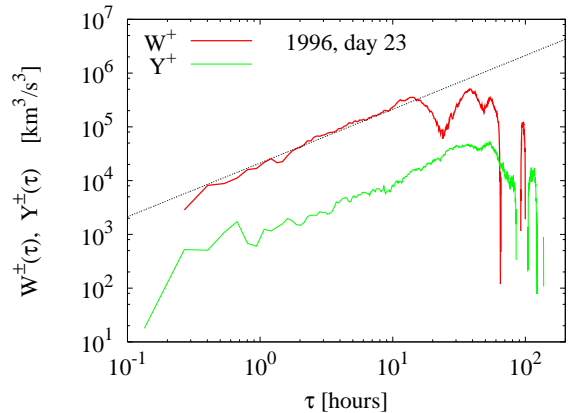}
 \caption{One example of the mixed third order compressible pseudo-energy
flux~$W^{+}(\tau)$ as computed from the Ulysses data during days~23 to~32
of~1996. The incompressible flux~$Y^{+}(\tau)$ in the same time window and a
linear fit are also indicated. In this case, both compressible and
incompressible fluxes obey a Yaglom-like law.}
 \label{fig1}
\end{figure}
\begin{figure}
 \includegraphics[width=8cm]{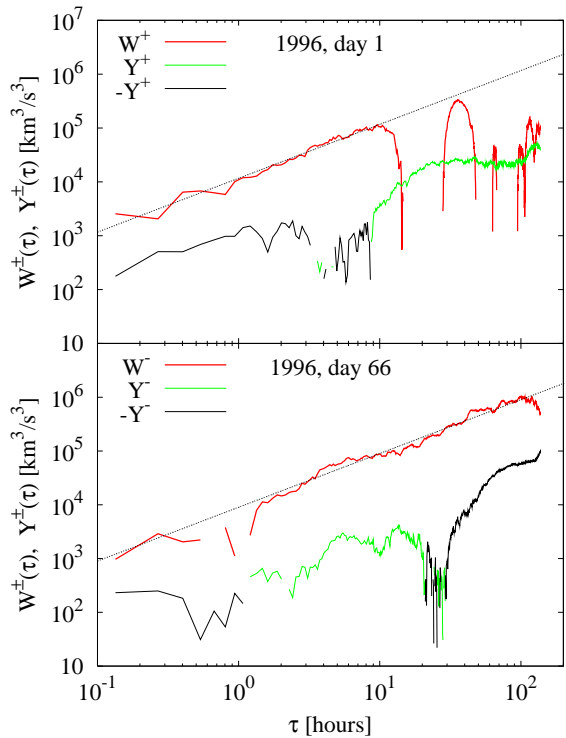}
 \caption{Top panel: an example of the third order compressible pseudo-energy flux~$W^{+}(\tau)$ 
during days~1 to~10 of~1996. Bottom panel: $W^{-}(\tau)$ for days~66 to~75 of the same
year. In both panels, the corresponding incompressible fluxes~$Y^{\pm}(\tau)$ (no scaling present) and a linear fit are displayed.}
 \label{fig2}
\end{figure}
Figure~\ref{fig1} shows one example of both mixed third-order structure
functions~$W^{+}(\tau)$ and~$Y^+(\tau)$ computed in the same
11~days windows where the scaling was observed.
Figure~\ref{fig2} shows two more examples of scaling, observed both
for~$W^{+}(\tau)$ and $W^{-}(\tau)$, in two different time windows. The
$W^{+}(\tau)$ scaling extends over 2~decades, while $W^{-}(\tau)$ behaves
linearly on the whole range of scales considered here (3~decades). In the last
example, the scaling is not present for the incompressible
fluxes~$Y^{\pm}(\tau)$. This example shows that the inclusion of compressible effect through
the density-weigthed fluctuations improves the scaling~(\ref{yagc}) and modify
the energy cascade.
The scaling relation~(\ref{yagc}) also allows a direct estimate of the
pseudo-energy transfer rates in the compressible case. 
A fit of the linear law~(\ref{scal}) provides the local values of the amount of
pseudo-energy transferred from large to small scales by the turbulent
MHD~cascade. This was already measured in the incompressive
case~\cite{noiprl,apj}, so that it is possible to compare the transfer rates in
the two cascades. The mean values, computed over the 46~observed scaling cases at
different radial distances from the sun,
($\pm$ their dispersion, in [J kg$^{-1}$ sec$^{-1}$]) for the compressible
cascade are $\epsilon^+ = 3668 \pm 1900$ (29~cases) and $\epsilon^- = 3536 \pm
2500$ (17~cases). Both values are considerably larger than the corresponding
values for the incompressive case ($\epsilon_I^+ = 182 \pm 73$, 24~cases, and
$\epsilon_I^- = 156 \pm 50$, 11~cases~\cite{apj}). This result shows again that the
cascade in the solar wind is strongly enhanced by density fluctuations, 
despite their small amplitude. Note that the new variables are built by coupling the
Els\"asser fields with the density, before computing the scale dependent increments. 
Moreover, the third order moments are very sensitive to intense field fluctuations 
(intermittency), that could arise when density fluctuations are correlated with 
velocity and magnetic field. 
Similar results, but with considerably smaller effect, were found in numerical 
simulations of compressive MHD~\cite{maclow}. We should point out that experimental 
values of energy transfer rate in the incompressive case had been also 
estimated with different tecniques from different datasets~\cite{vasquez,mc}. 
Those values are not in agreement with the present (incompressive case) results. 
However, the different nature of wind (ecliptic {\it vs} polar, fast {\it vs} slow, 
at different radial distances from the sun) makes such comparison only indicative.

An interesting open question is the problem of the solar wind heating. The
first models of solar wind assumed an adiabatic cooling due to spherical expansion of 
plasma blowing out of the sun. This would result in a radial decrease of the proton 
temperature $T(r) \sim r^{-\xi}$ with~$\xi = 4/3$. On the contrary, spacecraft
measurements~\cite{heating} have shown that the temperature decay is slower
than the adiabatic prescription, with~$\xi \in [0.7, 1]$. This implies that
some local heating mechanism is present. One standing
hypothesis is that the heating could be provided by energy dissipation
occurring at the small scales of a turbulent cascade~\cite{smith,vasquez,verma}.
Using equation~(\ref{yaglom}), the rates at which the incompressible turbulent
pseudo-energy is transported down the scales, and eventually dissipated at
small scale, can be measured directly from data. This has recently be used to
investigate whether or not a turbulent cascade can heat the solar wind. Results
were however not conclusive. In fact, the measured incompressible dissipation
rate of pseudo-energies can only account for up to~50\,\% of the solar wind
heating~\cite{mc,apj}.

\begin{figure}
 \includegraphics[width=8cm]{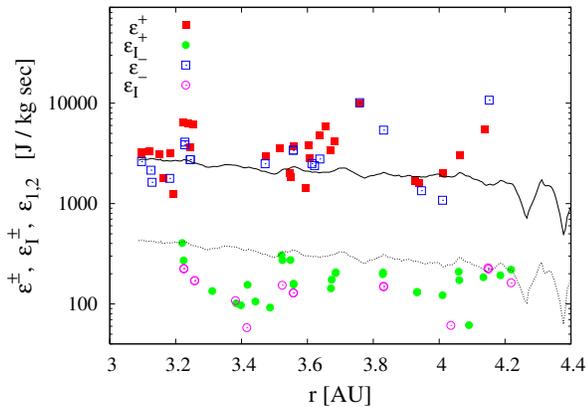}
 \caption{Radial profile of the pseudo-energy transfer rates obtained from the
turbulent cascade rate through the Yaglom relation, for both the compressive
and incompressive case. The solid lines represent the radial profiles of the
heating rate required to obtain the observed temperature profile.}
 \label{fig-heating}
\end{figure}
Figure~\ref{fig-heating} shows the radial profiles of the pseudo-energy
transfer rates for both the compressive and incompressive cascades. In the same
figure, we show the profiles of the heating rates needed to obtain the observed
temperatures, as estimated from heating models~\cite{vasquez,verma,apj} and
from the measured temperatures (the two different values refer to the
different estimates of the temperature obtained from Ulysses instruments). It
is evident that, while the incompressive cascade cannot provide all the energy
needed to heat the wind, the density fluctuations coupled with
magnetohydrodynamic turbulence can supply the amount of energy required. This
evidence shows the importance of the density fluctuations in polar, fast solar
wind turbulence, confirming that it should be considered as an example of
compressive fully developed MHD~turbulence. 
Note that, since in a few samples we measured both $\epsilon^+$ and $\epsilon^-$ 
in the same period, the values of the energy $\epsilon=(\epsilon^+ + \epsilon^-)/2$ 
and cross-helicity $\epsilon_H=(\epsilon^+ - \epsilon^-)/2$ transfer rates can be disentangled.
From the values obtained, it is clear that the cross-helicity contribution,
indicating the importance of the Alfv\'enic state of turbulence, can vary
from a negligible fraction (less than~1\,\%) to a considerable~25\,\% of the
energy contribution. Since its amplitude does not appear to be correlated with
the observation of the cascade, Alfv\'enicity seems not to play a crucial role
in the cascade at the observed scales. This would be in agreement with previous
analysis of solar wind turbulence anisotropy, where the Alfv\'enic contribution
to the field fluctuations is small~\cite{bieber,tim}.


To summarize, we used the density-weighted Els\"asser fields $\bw^{\pm}$ to
show for the first time that a phenomenological compressive Yaglom-like relation 
is verified to a large extent within the solar wind turbulence. 
This implies that low amplitude density fluctuations play a crucial role for 
scaling laws of solar wind turbulence~\cite{maclow}. 
This observation also confirm the recent results for the Kolmogorov $4/5$-law from numerical simulations of
compressible turbulence~\cite{apj1}, while no experimental evidences from real
fluids had been found so far.  This could be attributed to the incompressible
nature of flows in ordinary fluids accessible to laboratory experiments. Here
in fact, we present the first experimental observation of relation~(\ref{yagc}) in real systems.
Using solar wind data, we have access to a sample of weakly compressible MHD~turbulence in nature. 
Scaling law is found to be quite common and extends on a large range of scales, 
indicating not only that a nonlinear MHD cascade for pseudo-energies is active in the 
solar wind turbulence, but also that compressible effects are an important ingredient 
of the cascade. We point out that the observed departures from the scaling law could be due to 
presence of inhomogeneity and anisotropy in the solar wind~\cite{podesta2}.
The compressive corrections to the cascade also cause the transfer of a
considerably larger amount of energy toward the small scales, where it can be
dissipated to heat the plasma locally. The role of anisotropy in the solar wind turbulent  
cascade still remains an open question.




\begin{thebibliography}{99}

\bibitem{generale}
A.J. Hundhausen, Coronal Expansion and Solar Wind, Springer, New York (1972)

\bibitem{tu}
C.-Y. Tu and E. Marsch, Space Sci. Rev. \textbf{73}, 1 (1995)

\bibitem{noi}
R. Bruno and V. Carbone, Living Rev. Solar Phys. \textbf{2}, 4 (2005)

\bibitem{dmv}
M. Dobrowolny, A. Mangeney and P. Veltri, Phys. Rev. Lett. \textbf{45}, 144 (1980)

\bibitem{belcher}
J.M. Belcher and L. Davis Jr, J. Geophys. Res. \textbf{76}, 3534 (1971)

\bibitem{bruno1}
R. Bruno, B. Bavassano and U. Villante, J. Geophys. Res. \textbf{90}, 4373 (1985)

\bibitem{pp}
H. Politano, and A. Pouquet, J. Geophys. Lett. \textbf{25}, 273 (1998)

\bibitem{noiprl}
L. Sorriso-Valvo, et al., Phys. Rev. Lett. \textbf{99}, 115001 (2007)

\bibitem{vasquez}
B.J. Vasquez, et al. J. Geophys. Res. \textbf{112}, A07101 (2007)

\bibitem{mc}
B.T. MacBride, C.W. Smith, and M.A. Forman, Astrophys J. \textbf{679}, 1644 (2008)

\bibitem{nearly}
D. Montgomery, M.R. Brown, and W.H. Matthaeus, J. Geophys. Res. \textbf{92}, 282 (1987); W.H. Matthaeus, and M.R. Brown, Phys. Fluids \textbf{31}, 3634 (1988); G.P. Zank, and W.H. Matthaeus, Phys. Fluids \textbf{3}, 69 (1991); G.P. Zank, and W.H. Matthaeus, Phys. Fluids \textbf{5}, 257 (1993)

\bibitem{controversia}
W.H. Matthaeus, et al., J. Geophys. Res. \textbf{96}, 5421 (1991); B. Bavassano, and R.Bruno, J. Geophys. Res. \textbf{100}, 9475 (1995)

\bibitem{tm94}
C.-Y. Tu, and E. Marsch, J. Geophys. Res. \textbf{99}, 21481 (1994)

\bibitem{goldreich}
P. Goldreich, and S. Sridhar, Astrophys. J. \textbf{438}, 763 (1995)

\bibitem{cho}
J. Cho, and A. Lazarian, Phys. Rev. Lett.  \textbf{88}, 245001 (2002)

\bibitem{ligh55}
M.J. Lighthill, in IAU Symp. 2, Gas Dynamics of Cosmic Clouds (Amsterdam: North Holland), 121 (1955)

\bibitem{apj1}
A.G. Kritsuk, et al., Astrophys. J. \textbf{665}, 416 (2007)

\bibitem{apj2}
G. Kowal, and A. Lazarian, Astrophys. J. \textbf{666}, L69 (2007)

\bibitem{podesta1}
J. J. Podesta, et al., Nonlin. Process. in Geophys. \textbf{16}, 99 (2009)

\bibitem{dewit}
T. Dudok de Wit, Phys. Rev. E \textbf{70}, 055302R (2004)

\bibitem{apj}
R. Marino, et al., Astrophys J. \textbf{677}, L71 (2008)

\bibitem{maclow}
M.-M. Mac Low, Astrophys. J. \textbf{524}, 169 (1999)

\bibitem{smith}
C. W. Smith, et al., J. Geophys. Res. \textbf{106}, 8253 (2001)

\bibitem{verma}
M.K. Verma, D.A. Roberts, and M.L. Goldstein, J. Geophys. Res. \textbf{100}, 19839 (1995)

\bibitem{bieber}
J.W. Bieber, W. Wanner, and W.H. Matthaeus, J. Geophys. Res. \textbf{101}, 2511 (1996)

\bibitem{tim}
T.S. Horbury, M.A. Forman, and S. Oughton, Plasma phys. Control. Fusion  \textbf{47}, B703 (2005)

\bibitem{heating}
B.E. Goldstein, et al., Astron. Astrophys. \textbf{316}, 296 (1996)

\bibitem{podesta2}
J. J. Podesta, M. A. Forman and C. W. Smith, Phys. Plasmas \textbf{14}, 092305 (2007)

\end{thebibliography}
\end{document}